\documentclass[aps, prb, reprint, superscriptaddress, preprintnumbers, amsmath, amssymb, floatfix]{revtex4-2}
\usepackage{graphicx,float}
\usepackage{color}
\usepackage[first]{draftcopy}
\usepackage{hyperref}
\begin{document} %\raggedbottom

%\preprint{v.6.4 DEM/OOB}
\title{\boldmath $^{19}$F NMR and defect spins in vacuum-annealed LaO$_{0.5}$F$_{0.5}$BiS$_2$}

\author{S. Yadav}
\altaffiliation[Present address: ]{New Mexico Institute of Mining and Technology, Socorro NM, 87801, USA}
\author{S. Delgado}
\altaffiliation[Present address: ]{Division of Physics, Mathematics and Astronomy, Caltech, Pasadena CA 91125, USA}
\author{O.~O. Bernal}
\email[ Corresponding Author: ]{obernal@calstatela.edu.}
\affiliation{Department of Physics and Astronomy, California State University, Los Angeles, California 90032, USA}
\author{D.~E. MacLaughlin}
\affiliation{Department of Physics and Astronomy, University of California, Riverside, California 92521, USA}
\author{Y.~Liu}
\author{D.~Jiang}
\author{O. Santana}
\author{A. Mushammel}
\altaffiliation[Present address: ]{Michelson Laboratories, Inc., Commerce, California 90040, USA}
\affiliation{Department of Chemistry and Biochemistry, California State University, Los Angeles, Los Angeles, California 90032, USA}
\author{Lei Shu}
\author{K.~Huang}
\altaffiliation[Present address: ]{Lawrence Livermore National Laboratory, Livermore, California 94550, USA.}
\affiliation{State Key Laboratory of Surface Physics, Department of Physics, Fudan University, Shanghai 200433, China}
\author{D.~Yazici}
\altaffiliation[Present address: ]{
Faculty of Applied Physics and Mathematics and Advanced Materials Centre, 
Gda\'{n}sk University of Technology, 
%ul. Narutowicza 11/12,
Gda\'{n}sk, 80-233, Poland.
}
\author{M.~B. Maple}
\affiliation{Department of Physics, University of California, San Diego, La
Jolla, CA 92093, USA} 
\date{August 13, 2024}

\begin{abstract} %revised
We report results of magnetization and $^{19}$F NMR measurements in the normal state of as-grown LaO$_{0.5}$F$_{0.5}$BiS$_2$. The magnetization is dominated by a temperature-independent diamagnetic component and a field- and temperature-dependent paramagnetic contribution~$M_\mu(H,T)$ from a ~$\sim$1000~ppm concentration of local moments, an order of magnitude higher than can be accounted for by measured rare-earth impurity concentrations. $M_\mu(H,T)$ can be fit by the Brillouin function~$B_J(x)$ or, perhaps more realistically, a two-level $\tanh(x)$ model for magnetic Bi $6p$ ions in defect crystal fields. Both fits require a phenomenological Curie-Weiss argument~$x = \mu_\mathrm{eff}H/(T + T_W)$, $T_W \approx 1.7$~K\@. There is no evidence for magnetic order down to 2~K, and the origin of $T_W$ is not clear. $^{19}$F frequency shifts, linewidths, and spin-lattice relaxation rates are consistent with purely dipolar $^{19}$F/defect-spin interactions. The defect-spin correlation time~$\tau_c(T)$ obtained from $^{19}$F spin-lattice relaxation rates obeys the Korringa relation~$\tau_cT = \text{const.}$, indicating the relaxation is dominated by conduction-band fluctuations.
\end{abstract}

\maketitle

\section{\label{sec:intro} INTRODUCTION} 

The discovery of layered BiS$_2$-based superconductors~\cite{MFGS12, MDDT12} opened a new venue of research in the field, adding considerably to the existing list of layered superconducting materials. In particular, the rare-earth-based compound series~$Ln$O$_{1-x}$F$_x$BiS$_2$, $Ln = \text{La}$, Ce, Pr, Nd, Sm, and Yb, has been studied extensively in the search for higher superconducting temperatures 
($T_c$\,s); see e.g.,~\cite{WWJYHM2013, YJWM15, SGAMKMTM2019} and references therein. 
The highest $T_c$ has been achieved for $Ln = \text{La}$, for which $T_c^\textrm{max} \sim 11.5$~K for samples annealed under hydrostatic pressure (HP)~\cite{M2019}. 
LaO$_{1-x}$F$_x$BiS$_2$ (LOFBS) grown at ambient pressure reaches $T_c \sim 10.6$~K~\cite{WYWHM2013, M2019} under pressure, which is the next highest value of $T_c$ in this series. 

As-grown LOFBS powder samples ($T_c \sim 3$~K) show signs of local superconducting phases for temperatures as high as 10~K~\cite{M2019}. This should be suggestive, and in fact similar signs observed in tunneling spectra for Bi$_4$O$_4$S$_3$~\cite{LYFWTDW2013} inspired the search for higher $T_c$ in the BiS$_2$ compounds~\cite{M2019}. 
Replacement of La by other Ln ions is seen to induce stable bulk SC as the ion's size decreases, but with $T_c$s less than that for La~\cite{YJWM15}. 
Similarly, stabilizing bulk superconductivity by replacing S for Se does not yield any higher $T_c$; see Ref.~\cite{M2019} for details. 

The reasons for the lack of coherence in the superconducting fluid have apparently not been elucidated. The importance of magnetism to superconducting behavior might also be expected by the prediction~\cite{ZLLFZ2014} and potential confirmation~\cite{WSMTYKUANWTO2017} of spatial spin textures induced by spin-orbit coupling via Rashba~\cite{BR1984} and Dresselhaus~\cite{D1955} effects in other systems. There appears to be a consensus regarding the conventional nature of the superconductivity in LOFBS.

Very little has been reported on the magnetic properties of the normal state from which the superconducting phase develops~\cite{M2015,M2017,M2019}. Magnetic susceptibility measurements reported to date do not go to temperatures higher than $\sim$20~K or applied fields higher than those for which the effects of magnetism can be avoided or minimized. No NMR measurements have been reported on this compound.

 %revised
Magnetization and $^{19}$F~NMR measurements have been carried out on as-grown vacuum-annealed LOFBS\@. The observed magnetization is diamagnetic and constant above 100~K, below which it exhibits Brillouin-function/Curie-Weiss-like field and temperature dependencies typical of local-moment paramagnetism. With a phenomenological Curie-Weiss form for the argument~$x = \mu_\mathrm{eff} H_0/k_B(T + T_W)$], either a Brillouin function~$B_J(x)$ or the two-level version~$\tanh(x)$ provides a good fit to the data with fit values~$\sim$1000~ppm for the local-moment concentration. However, measurement of impurity levels by mass spectrometry reveals impurity levels of 150~ppm Ce, 31~ppm~Gd, and traces of other $3d$ and $4f$ impurities. Thus the dilute local-moment magnetization in LOFBS is intrinsic, most likely due to Bi $6p$ ions decoupled from the conduction band by defects.

This is reminiscent of relatively recent findings in studies of $sp$ systems containing no $d$ or $f$ ions. Dilute magnetic moments of order~1~$\mu_B$ have been associated with defect-induced magnetism in graphite and graphene~\cite{KETM00, ESHSHB03, YH07, CKF09, SNRNTW10, SZK11}. Non-carbon materials have also been found to house magnetic moments of order 1 to 4$\mu_B$ induced by defects; these include wide-gap III nitrides (GaN and BN)~\cite{DXZ08} and the dielectric oxide system HfO$_2$~\cite{VFC04,PS05}. Disordered ZnO and TiO$_2$ are also magnetic~\cite{PYSWYL07,EHSB20}. 

Pertinent to the present study, Bi-containing materials have also been found to display unexpected magnetism~\cite{AVKOGFB94, KNKO96, KOS06}. These systems exhibit weak paramagnetism, and local fields of up to 250~G are observed at Bi sites using NQR, SQUID and $\mu$SR techniques. A strong coupling between CEF and magnetic order is reported for Bi-based oxides in which a magneto-electric effect linear in magnetic field is observed~\cite{KOS06}. Reviews~\cite{EHSS13, EHSB20} give further information and references on this topic.

 %revised
$^{19}$F NMR frequency shifts, linewidths, and spin-lattice relaxation rates have been measured. Spin echos~\cite{Slic96} were used to determine the shift and line shape (the Fourier transform of the echo shape) due to the static field distribution at $^{19}$F sites. The $^{19}$F line in LOFBS is Voigtian, i.e., a convolution of Gaussian and Lorentzian distributions. The Lorentzian linewidth~$1/T^*_{2e}(T)$ is expected from dipolar and/or RKKY coupling to dilute local moments~\cite{[{See, e.g., }] [{ and references therein.}] WaWa74}; our observation that the NMR frequency shift~$K(T)$ and $1/T^*_{2e}(T)$ both vary linearly with the intrinsic magnetization~$M_\mu(T)$\footnote{Or, equivalently, with the intrinsic susceptibility~$\chi_\mu(T) = M_\mu(T)/H_0$.} is good evidence for this and their common origin. 

The observed $^{19}$F NMR spin-lattice relaxation function is a stretched exponential~$\exp[-(t/\tau_1)^p]$, $p < 1$. This indicates a spatially inhomogeneous distribution of relaxation rates, as expected from a dilute local-moment scenario~\cite{MSW71, *MSW72, TsHa68, John06}. The relaxation rate~$1/\tau_1(T)$ exhibits a non-monotonic temperature dependence with a maximum in the neighborhood of 10~K, and is considerably suppressed by field. The power~$p$ is roughly constant (0.70--0.75) at low temperatures, increasing slightly above 100~K\@. The observed values of $1/T^*_{2e}$ and $1/\tau_1(T)$ are consistent with a dipolar $^{19}$F/defect-spin coupling, and rule out a significant conduction-band-mediated RKKY interaction between defect spins and $^{19}$F nuclei. The $1/\tau_1(T)$ data also place a low upper limit on any $^{19}$F Korringa relaxation rate due to conduction-band spin fluctuations. 

 %revised
Thus the $^{19}$F relaxation is dominated by fluctuations of the defect spins. These exhibit a temperature-dependent correlation time~$\tau_c(T)$ roughly of the Korringa form~$\tau_c(T)T = \text{const.}$, indicating a dominant defect-spin/exchange coupling with conduction-band states. This would be inconsistent with extrinsic magnetic impurities replacing La$^{3+}$ in the blocking layers of the crystal structure, as would be the case for Ce or Gd ions, because of the considerable distance ($>$6~\AA) between blocking and conduction layers. The results suggest instead a small but significant exchange interaction~${\cal J}_\mathrm{ex} = 3.6$~meV between defect spins and conduction-layer band states, consistent with the defects being associated with BiS$_2$ planes.

Our measurements of magnetization and $^{19}$F NMR in LOFBS lead us to conclude that the observed local-moment magnetism is due to a dilute distribution of crystal defects. This may be related to recent results~\cite{AL2019} where bond disorder in the BiS$_2$ planes is described as potentially being a conduction-electron trap. Our results indicate that the magnetic defects interact principally with the BiS$_2$-plane conduction band rather than each other, but that ${\cal J}_\mathrm{ex}$ is far too weak to have a significant effect on the superconducting state.

The article is organized as follows. Section~\ref{sec:expt} gives brief descriptions of sample preparation and experimental techniques. Results of magnetization and $^{19}$F NMR lineshape, linewidth, frequency shift, and spin-lattice relaxation measurements are reported in Sec.~\ref{sec:results} and discussed in Sec.~\ref{sec:disc}. A summary and conclusions follow in Sec.~\ref{sec:concl}.

\section{\label{sec:expt} EXPERIMENT} 

\subsection{\label{exp:sample} Sample} 

Polycrystalline LOFBS was prepared by solid-state reaction at ambient pressure as 
described previously~\cite{YHWCFM2013,ZHDM16}. 
The sample was characterized by 
X-ray powder diffraction, electrical resistivity, magnetic susceptibility and 
specific heat measurements to determine its superconducting properties. 
The starting purity of the precursor powders of La$_2$O$_3$, LaF$_3$, 
La$_2$S$_3$ and Bi$_2$S$_3$ used in the preparation of the sample exceeded 
99.9\% in most cases~\cite{YHWCFM2013}. 

% revised
Trace magnetic impurities in the sample were fully characterized. Preliminary analytical studies indicated Ce and Gd as dominant impurities in the starting material. A Perkin-Elmer ELAN 9000 inductively-coupled plasma mass spectrometer (ICP-MS) was used to determine impurity concentrations. Initial measurements without reference standards yielded Ce and Gd concentrations of $\sim$90 and $\sim$17~ppm, respectively, with Sm about the same as Gd, Tb and Yb considerably lower ($\lesssim$5~ppm) and Nd, Dy, Ho, Tm, Co, Cr, Fe, and Ni all $\lesssim$1~ppm.

% revised
With reference standards the accuracy of an ICP-MS measurement is improved considerably. Ce and Gd standard solutions with concentrations 1--200~ppb were prepared from commercial ICP-MS standards (1000~ppm) through serial dilution using using 2\% nitric acid. The standard solutions were then used to create calibration curves for Ce and Gd. To prepare the LOFBS ICP-MS samples, 29.68~mg of material was first fully digested in 5~mL %trace metal
nitric acid in a heated sonicator. The digested sample solution was then diluted to 250~mL in a volumetric flask using 2\% nitric acid. Three pieces were measured from different parts of the sample for consistency. The results were 150(3)~ppm Ce and 31(1)~ppm Gd, where the uncertainties are the variations between the pieces. 

\begin{figure} [ht]
\includegraphics[clip = ,width = 0.3\textwidth]{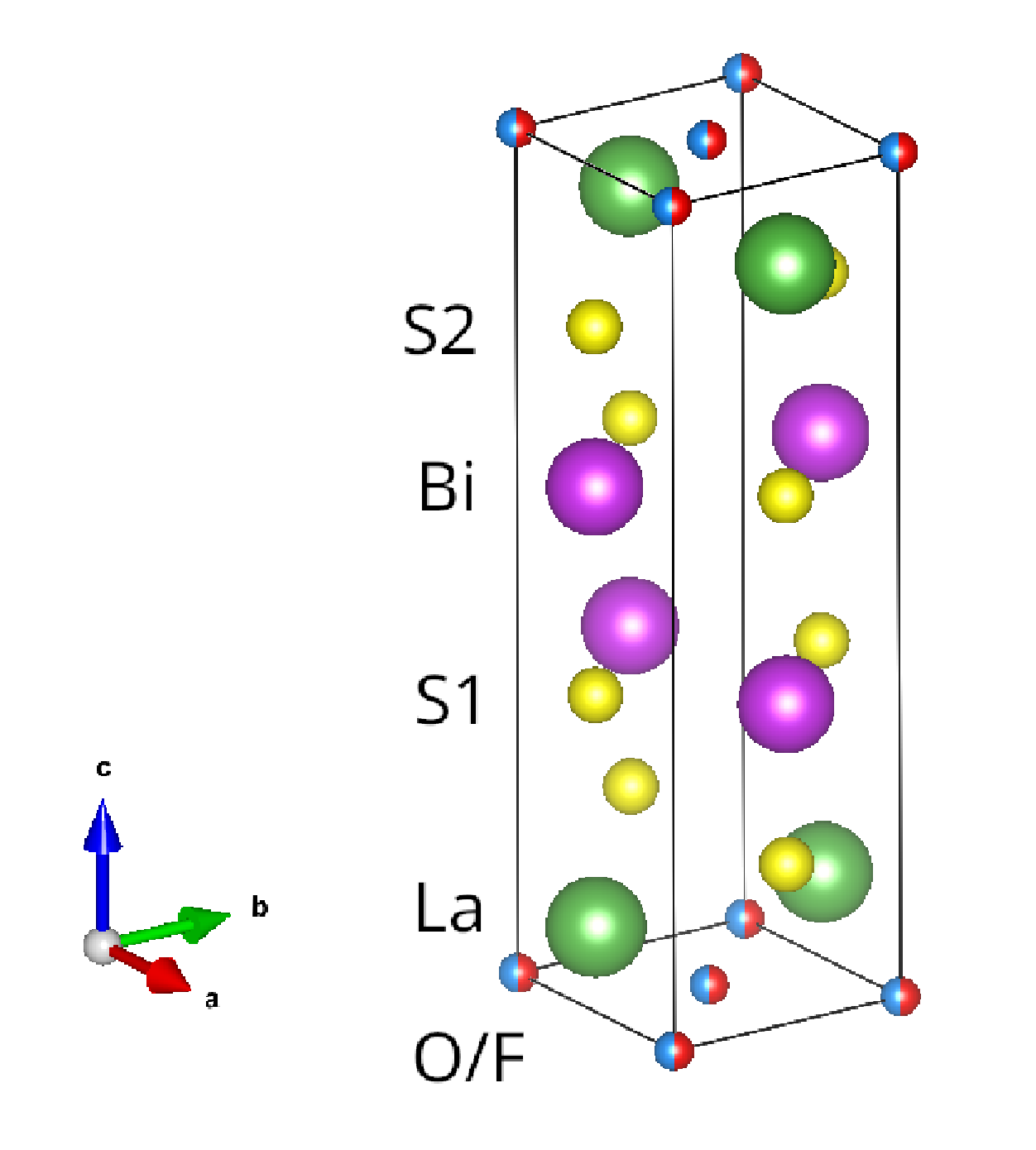}
\caption{\label{fig:struct} Unit cell of LOFBS.}
\end{figure}

The unit cell of LOFBS is shown in Fig.~\ref{fig:struct}. 
Conduction-band states and superconductivity are considered to be restricted to the Bi-S bilayers; the blocking La-O/F bilayers are insulating.

\subsection{\label{exp:mag} Magnetization} 

The sample magnetization was measured using a Quantum Design PPMS system configured with a vibrating sample magnetometer. Measurements were made for temperatures between 1.8~K and 300~K and magnetic fields in the range 0.5--80~kOe. 

% subsection revised
\subsection{\label{exp:NMR} NMR} 

$^{19}$F is a spin-1/2 nucleus with a large %nuclear
magnetic moment ($2.63\mu_N$). It is not affected by electric field gradients, and the resonance line in LOFBS is narrow (Sec.~\ref{rslts:spectra}). $^{19}$F NMR experiments were carried out using a standard pulsed NMR spectrometer with phase-sensitive detection over the temperature range 1.8--300~K for applied fields of approximately 12, 20, and 40~kOe. $^{19}$F Spin echoes were produced using the Hahn $\pi/2$-$\pi$ pulse sequence~\cite{[{See, e.g., }] Slic96}. 

\paragraph{\label{exp:freqshift} Frequency shifts.} $^{19}$F frequency shifts were measured in an applied field~$H_0 = 11.741$~kOe for temperatures between 1.8 and 300~K\@. The $^{19}$F frequency in CaF$_2$~\cite{Bruc57, LoNo57, BL1966} was used as a reference; samples of both LOFBS and CaF$_2$ were included in the NMR coil~\cite{BL1966}. A careful characterization of the CaF$_2$ line was carried out to confirm it provided a consistent reference over the full range of temperatures. This line is nearly a rectangle in the frequency domain~\cite{Bruc57}, so that the time-domain signals exhibit Lowe-Norberg beats~\cite{LoNo57}. To facilitate the fits the CaF$_2$ lineshape was approximated by a power exponential function~$f(\nu) = \exp\{-[(\nu{-}\nu_0)/\Delta\nu)^q\}$ with $q \approx 5$ from the fits; this shape is independent of temperature. Fits to the combined spectra are described below in Sec.~\ref{rslts:freqshift}. 

\paragraph{\label{exp:linewidth} Line shapes, linewidths.} Line shapes and linewidths were determined from relaxation rates derived from the spin-echo shapes (Sec.~\ref{rslts:echo}). These were fit to the product of a Gaussian~$\exp\left[\frac{1}{2}(t/T^*_{2G})^2\right]$ and a Lorentzian~$\exp\left(t/T^*_{2e}\right)$, so that the relaxation rates are the linewidths of the corresponding components of the Voigtian. In LOFBS these are both considerably faster than dynamic spin-lattice relaxation rates (Sec.~\ref{rslts:slrelax}), and are therefore attributed to broadening by inhomogeneous distributions of static local fields. 

\paragraph{\label{exp:SLR} Spin-lattice relaxation.} The narrow line allowed complete inversion of the $^{19}$F magnetization by a single rf pulse. The recovery signal~$S(t)$ after this pulse is related to the normalized relaxation function~$s(t)$ [$s(0) = 1$, $s(\infty) = 0$] by
\begin{equation} \label{eq:rlxsig}
S(t) = S_f + (S_i - S_f)s(t) \,,
\end{equation}
where $S_i$ and $S_f$ are the initial and final recovery signals respectively. Results are discussed below in Sec.~\ref{rslts:slrelax}.

\section{\label{sec:results} RESULTS} 

\subsection{\label{rslts:magresults} Magnetization} 

\begin{figure} [ht]
\includegraphics[clip = ,width = 0.4\textwidth]{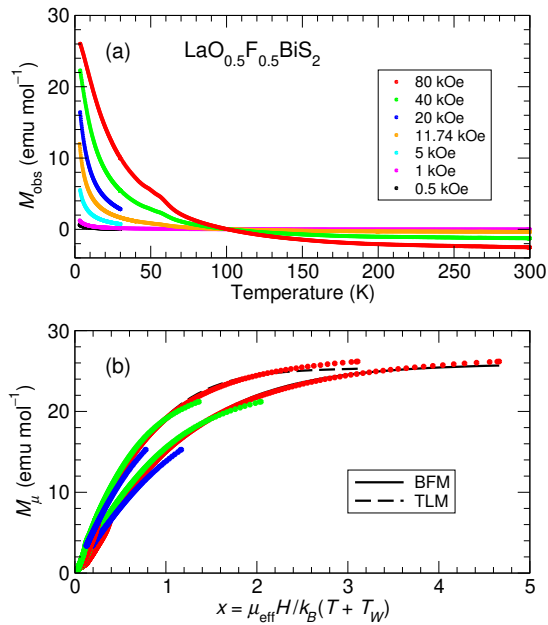}
\caption{\label{fig:parafit} (a)~Temperature dependence of observed magnetization $M_\mathrm{obs}(H,T)$ in LOFBS powder. (b)~Fits of the Brillouin function~$B_J(x)$ (solid curve) and $\tanh(x)$ (dashed curve), $x = \mu_\mathrm{eff}H/k_B(T+T_W$), to the intrinsic 
magnetization~$M_\mu(H,T)$. See text and Sec.~\ref{disc:Tdepmag} for details.}
\end{figure}

Figure~\ref{fig:parafit}(a) shows the observed magnetization $M_\mathrm{obs}(H,T)$ in LOFBS\@. 
The sample is diamagnetic above $\sim$100~K, and exhibits the Curie-Weiss-like behavior typical of local-moment paramagnetism. The temperature-dependent paramagnetism~$M_\mu(H,T)$ was obtained by subtraction of a temperature-independent diamagnetic term~$M_0(H)$ and small contributions from the trace Ce and Gd impurities (Sec.~\ref{exp:sample} above). 

% revised
Two local-moment models have been fit to the $M_\mu(H,T)$ data, as shown in Fig.~\ref{fig:parafit}(b). The Brillouin function model (BFM, solid curve) uses a modified Brillouin function~\cite{[{See, e.g., }] [{, p.\ 655.}] AsMe76} 
\begin{eqnarray} \label{eq:brillfit}
M_\mu(H,T) & = & N_0c\mu_\mathrm{eff}B_J(x) \,, \nonumber \\
x & = & \frac{\mu_\mathrm{eff}H}{k_B(T+T_W)} \,, \quad \text{(BFM)}
\end{eqnarray}
where $N_0$ is the formula-unit density, $c$ is the concentration of local moments, and $\mu_\mathrm{eff}$ is the effective local magnetic moment. The Weiss temperature~$T_W$ is included phenomenologically because of the observed Curie-Weiss behavior of the low-field susceptibility. As discussed in Sec.~\ref{disc:Tdepmag}, this is not a good approximation at high fields, which is likely to account for the deviations for large $x$.

% revised
The two-level model (TLM, dashed curve) uses a hyperbolic tangent function
\begin{equation} \label{eq:twolevelfit}
M_\mu(H,T) = cN_0\mu_\mathrm{eff}\tanh(x) \,, \quad \text{(TLM)}
\end{equation}
with $x$ as in Eq.~(\ref{eq:brillfit}). This is the BFM with $J$ fixed at 1/2, but the interpretation of the effective moment that results from the model differs in the two models as discussed below in Sec.~\ref{disc:Tdepmag}. 

% revised
The fits and resulting parameter values are discussed in Sec.~\ref{disc:Tdepmag}. $M_\mu$ is consistent with $\gtrsim$1100~ppm of local moments, an order of magnitude or more larger than the measured trace Gd and Ce levels. The large-$x$ deviations are smaller for the BFM than for the TLM, but this cannot be taken as evidence for the former because of the inapplicability of the Curie-Weiss approximation noted above.

\subsection{\label{rslts:spectra} \boldmath $^{19}$F Spectra} 

\subsubsection{\label{rslts:freqshift} \boldmath $^{19}$F frequency shifts} 

\begin{figure} [ht]
\includegraphics[clip = ,width = 0.4\textwidth]{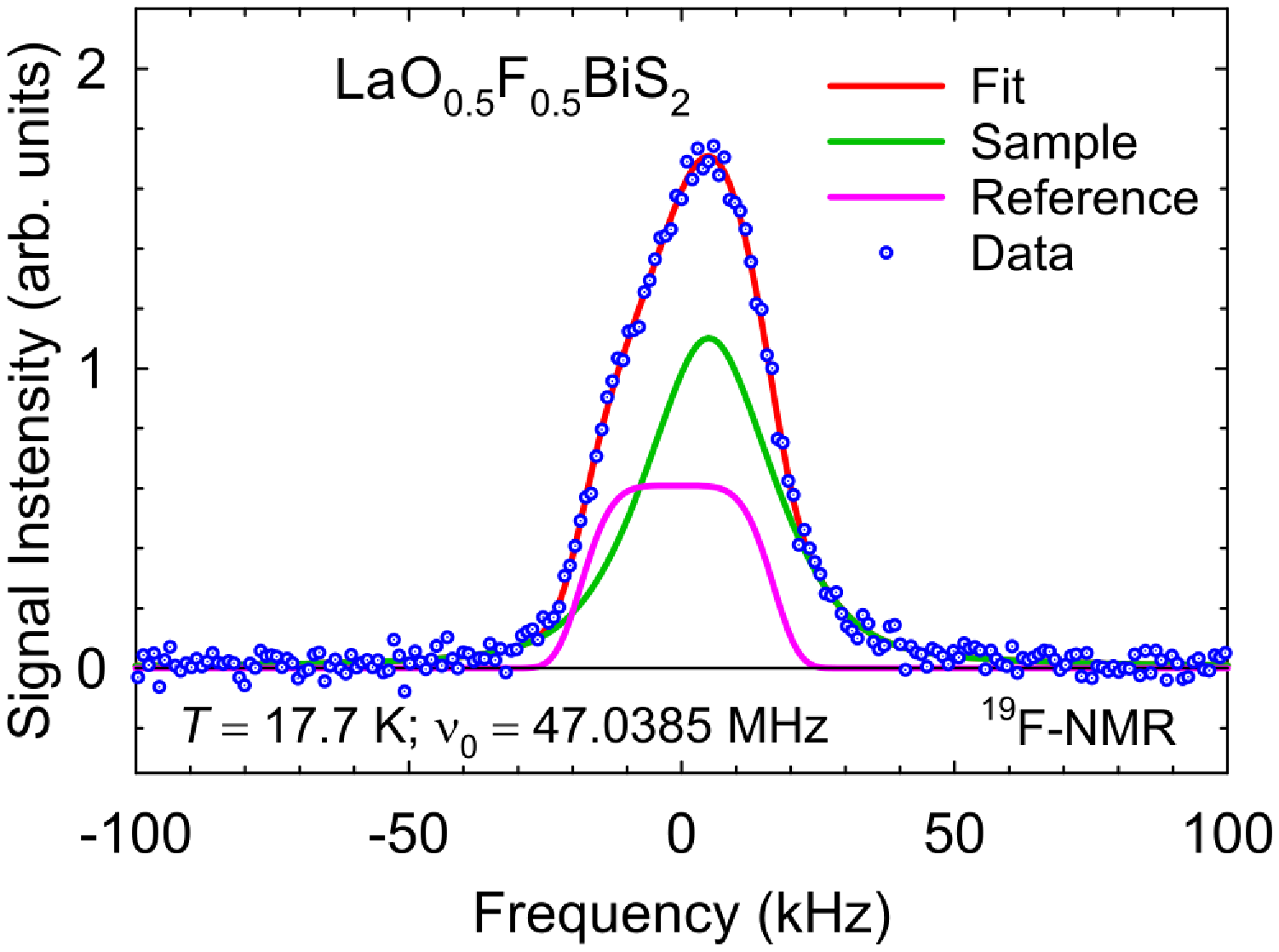}
\caption{\label{fig:K-Fit} Fit of a two-component line to a $^{19}$F spectrum from a sample containing both LOFBS powder and a CaF$_2$ reference.}
\end{figure}

Frequency shifts were obtained from Fourier transforms of spin echos from samples containing both LOFBS and CaF$_2$. An example of the fit of an observed spectrum to the sum of these contributions is shown in 
Fig.~\ref{fig:K-Fit}. 
The center of the LOFBS line with respect to that of CaF$_2$ could be determined by fitting the combined spectra to a sum of two functions: a modified Gaussian for CaF$_2$ (Sec.~\ref{exp:freqshift}) and a symmetric function for LOFBS\@. For the latter Gaussian and Lorentzian fits yield the same line centers to within statistical errors, but a pseudo-Voigtian~\cite{[{See, e.g., }] CrCy97} superposition of Gaussian and Lorentzian terms accounts for the spectral tails better and reduces the systematic uncertainty of the shift. 

\begin{figure}[ht]
\includegraphics[clip = ,width = 0.4\textwidth]{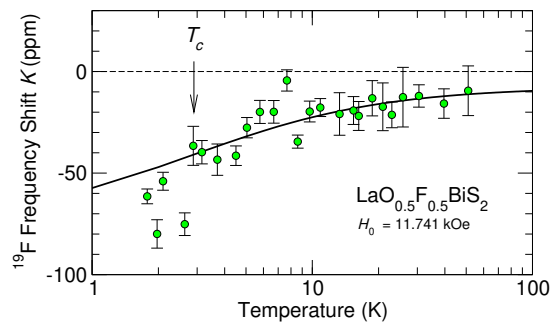}
\caption{\label{fig:KvsT} Temperature dependence of $^{19}$F frequency shift in LOFBS powder with respect to a CFCl$_3$ reference. Curve: $K(T)$ from a linear Clogston-Jaccarino fit to $K$-$\chi_\mu$ data (Fig.~\ref{fig:Kvschi}).}
\end{figure}

Figure~\ref{fig:KvsT} gives the temperature dependence of the frequency shift~$K$ with respect to CFCl$_3$, a reference material for which the field at the $^{19}$F site is close to the vacuum value~\cite{SBLB11}. 
The CaF$_2$ chemical shift of $-108.0(2)$~ppm with respect to CFCl$_3$~\cite{SBLB11} was used to correct the observed shift. For $T > T_c = 2.9$~K~\cite{ZHDM16} $|K(T)|$ decreases with increasing temperature, reminiscent of $1/T^*_{2e} (T)$ (Fig.~\ref{fig:echorates} below). It is well fit by a linear Clogston-Jaccarino~\cite{CGJY64} plot~$K(T) = K_0 + A_K\chi_\mu(T)$ (Fig.~\ref{fig:Kvschi}), where $\chi_\mu$ is the defect susceptibility; this yields the curve in Fig.~\ref{fig:KvsT}. For $T < T_c$ the data are scattered but lie somewhat below the normal-state fit curve. 

\subsubsection{\label{rslts:echo} Spin-echo shapes, linewidths}

\begin{figure} [ht]
\includegraphics[clip = ,width = 0.4\textwidth]{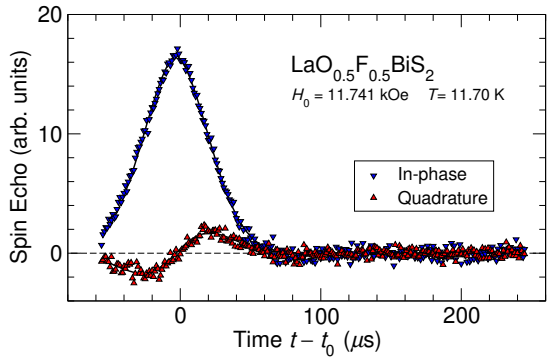}
\caption{\label{fig:echo} Representative $^{19}$F in-phase and quadrature spin-echo signals in LOFBS powder. Curves: fits of Eqs.~(\ref{eq:echoi}) and (\ref{eq:echoq}) to the data.}
\end{figure}

For the $^{19}$F relaxation measurements a sample containing only LOFBS (no CaF$_2$) was prepared. Figure~\ref{fig:echo} shows typical in-phase and quadrature spin-echo signals. 

\noindent Voigtian functions
\begin{eqnarray} \label{eq:echoi}
S_{ip}(t) & = & \exp\{-|t-t_0|/T^*_{2e}-\textstyle{\frac{1}{2}}[(t-t_0)/T^*_{2G}]^2\} \nonumber \\
& & {\times} \cos[\delta\omega(t-t_0)] 
\end{eqnarray}
and
\begin{eqnarray} \label{eq:echoq}
S_q(t) & = & \exp\{-|t-t_0|/T^*_{2e}-\textstyle{\frac{1}{2}}[(t-t_0)/T^*_{2G}]^2\} \nonumber \\
& & {\times} \sin[\delta\omega(t-t_0)]
\end{eqnarray}
were fit to the in-phase and quadrature echo signals, respectively. Here $t_0$ is the refocusing time of the echos, and the cosine and sine factors account for any difference~$\delta\omega$ between the spectrometer and $^{19}$F frequencies. 

Phase-sensitive detection and simultaneous fit of both signals is essential: the Gaussian and cosine factors in $S_{ip}(t)$ are the same to second order and are therefore strongly correlated statistically, so that $T^*_{2G}$ and $\delta\omega$ cannot be determined separately from a fit of Eq.~(\ref{eq:echoi}) alone.

\begin{figure} [ht] 
\includegraphics[clip = ,width = 0.4\textwidth]{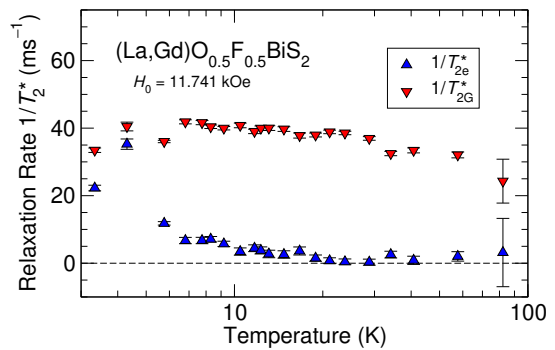}
\caption{\label{fig:echorates} Temperature dependence of static $^{19}$F relaxation rates~$1/T_2^*$ in LOFBS powder from spin-echo data. $1/T_{2\mathrm{e}}^*$: exponential rate. $1/T_{2\mathrm{G}}^*$: Gaussian rate.}
\end{figure}

Figure~\ref{fig:echorates} gives the temperature dependencies of $1/T^*_{2e}$ and $1/T^*_{2G}$. 
Gaussian relaxation is dominant, with a nearly temperature-independent rate of roughly $40~\text{ms}^{-1} \approx 1.6$~Oe in field units. Dipolar coupling to nearby $^{19}$F and $^{139}$La nuclei~\cite{[{See, e.g., }] [{, Chap.~IV(III).}] Abra61} yields a Gaussian width of $\sim 1.1$~Oe, in reasonable agreement. 

The exponential rate~$1/T^*_{2e}$ is too small to be measured accurately above $\sim$30~K, and increases at lower temperatures. A Lorentzian contribution to the static field distribution proportional to the magnetization is expected from a dilute concentration of paramagnetic local moments~\cite{[{See, e.g., }] [{ and references therein.}] WaWa74}. 

\subsection{\label{rslts:slrelax}\boldmath $^{19}$F Spin-Lattice Relaxation}

\begin{figure} [ht]
\includegraphics[clip=,width=0.4\textwidth]{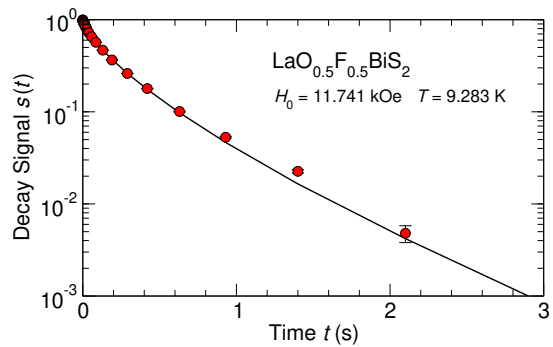}
\caption{\label{fig:relaxationsig} Semilogarithmic plot of observed $^{19}$F spin-lattice relaxation function~$s(t)$ in LOFBS powder, showing upward curvature characteristic of inhomogeneously distributed relaxation.}
\end{figure}

Figure~\ref{fig:relaxationsig} shows an example of an observed spin-lattice relaxation function~$s(t)$ [Eq.~(\ref{eq:rlxsig})] from LOFBS\@. The upward curvature on the semilogarithmic plot indicates an inhomogeneous distribution of relaxation rates~\cite{TsHa68, MSW71, MSW72, John06}. 
It is often modeled by a stretched exponential~\cite{[{See, e.g., }] John06}
\begin{equation} \label{eq:stretchexp}
s(t) = \exp[-(t/\tau_1)^p] \,, \quad p < 1 \,,
\end{equation}
motivated primarily by the result for direct host spin-lattice relaxation (no nuclear spin diffusion) by dilute paramagnetic local moments with a $1/r^3$ interaction. In a three-dimensional host $p = 1/2$~\cite{Blum1960, TsHa68, MSW72}; in a two-dimensional host $p = 1/3$~\footnote{Unpublished; the calculation of Ref.~\cite{MSW72} for three dimensions is easily modified for two dimensions.}. 

\begin{figure} [ht] 
\includegraphics[clip = ,width = 0.4\textwidth]{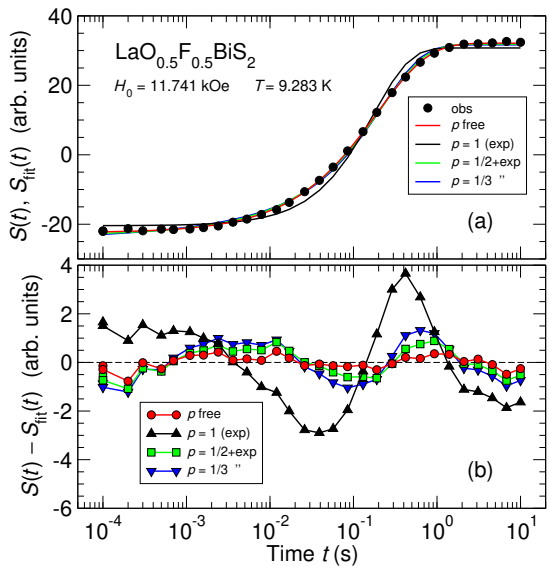}
\caption{\label{fig:diffplot} (a)~Recovery of $^{19}$F spin-lattice relaxation signal in LOFBS powder, $H_0= 11.74$~kOe, $T = 9.283$~K\@. Points: observed~$S(t)$. Curves: fit functions~$S_\mathrm{fit}(t)$ (see text). (b)~Differences~$S(t) - S_\mathrm{fit}(t)$.}
\end{figure} 
The recovery signal~$S(t)$ after pulse inversion from which $s(t)$ of Fig.~\ref{fig:relaxationsig} was derived is shown in Fig.~\ref{fig:diffplot}(a) (points), together with fits of various functions~$S_\mathrm{fit}(t)$ to the data. These are
\begin{enumerate}

\item $p$ free: a stretched exponential~$\exp[-(t/\tau_1)^p]$ with $p$ a fit parameter,

\item $p = 1$: a simple exponential~$\exp[-(t/T_1)]$,

\item $p = 1/2$: the product~$\exp[-(t/\tau_1)^{1/2}]\exp(-t/T_1)$ of a square-root exponential and a simple exponential, and

\item $p = 1/3$: the product~$\exp[-(t/\tau_1)^{1/3}]\exp(-t/T_1)$ of a cube-root exponential and a simple exponential.

\end{enumerate}
Figure~\ref{fig:diffplot}(b) gives the differences~$S(t) - S_\mathrm{fit}(t)$.

Fits to fixed-power stretched exponentials without the simple-exponential factor ($1/T_1 = 0$) are significantly worse. A fit to the product~$\exp[-(t/\tau_1)^p]\exp(-t/T_1)$ with $1/\tau_1$, $p$, and $1/T_1$ all free was not possible; the parameters are too strongly correlated. It can be seen from Fig.~\ref{fig:diffplot}(b) that a stretched exponential with $p$ a fitting parameter gives the best fit. 

\begin{figure} [ht]
\includegraphics[clip = ,width = 0.4\textwidth]{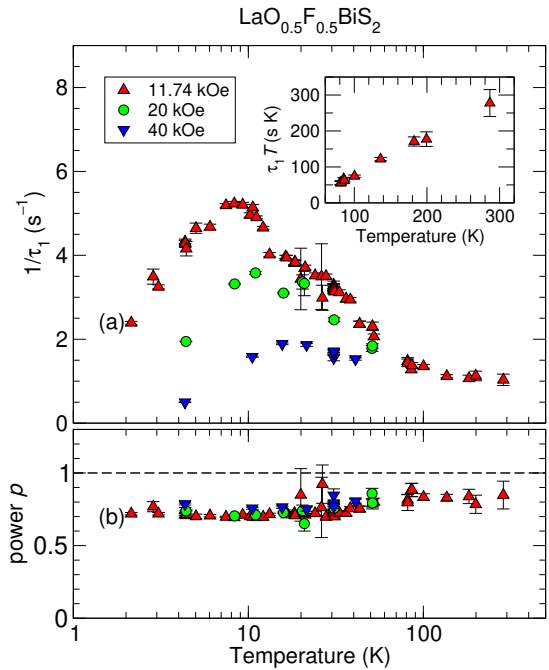}
\caption{\label{fig:pexprelax} $^{19}$F stretched-exponential relaxation in LOFBS powder. (a)~Stretched-exponential rate~$1/\tau_1(T,H)$ [Eq.~(\ref{eq:stretchexp})]. Inset: $\tau_1(T)T$, $H_0 = 11.74$~kOe, at high temperatures. (b)~Stretching power~$p(T,H)$.}
\end{figure}

Values of $1/\tau_1$ and $p$ for fits using function~No.~1 above are shown in Fig.~\ref{fig:pexprelax}. 
For each field $1/\tau_1(T)$ is strongly non-monotonic, with a maximum in the neighborhood of 10~K\@. The rate is considerably suppressed by field. The power~$p$ is roughly constant at 0.70--0.75 at low temperatures, increasing to $\sim$0.85 above 100~K\@. It exhibits little if any field dependence.

\section{\label{sec:disc} DISCUSSION}

\subsection{\label{disc:magnetiz} Magnetization}

\subsubsection{\label{disc:Tindepchi}Temperature-independent magnetization} 

The temperature-independent susceptibility~$\chi_0 = M_0(H)/H$ of LOFBS was obtained from plots of $M_\mathrm{obs}(H,T)$ vs $1/T$ by extrapolating to $1/T = 0$. $\chi_0 = \chi_\mathrm{core}^\mathrm{dia} + \chi_P + \chi_L + \chi_\mathrm{orb}$, where $\chi_\mathrm{core}^\mathrm{dia}$ is the core diamagnetic susceptibility~\cite{MBM70, CBK77}, and $\chi_P$, $\chi_L$, and $\chi_\mathrm{orb}$ are the Pauli (paramagnetic), Landau (diamagnetic), and orbital contributions, respectively. 

$\chi_P$ was estimated from reported Sommerfeld specific-heat coefficients~$\gamma_\mathrm{spht} = C_\mathrm{el}(T)/T$~\cite{YHWCFM2013, SLYZ14, KTNT2017} using the free-electron relation~$\chi_P = 3(\mu_B/\pi k_B)^2 \gamma_\mathrm{spht}$. This includes correlation effects but does not take exchange enhancement into account~\cite{CBK77}. The Landau and Pauli susceptibilities are related: $\chi_L = -\frac{1}{3}\chi_P(m/m^*)^2$~\cite{[{See, e.g., }] AsMe76}, where $m$ is the electron mass and $m^*$ is the effective mass including correlation effects. Estimates of $m/m^*$ have been obtained~\cite{Baty09} from the ratios of the above $\chi_P$ estimates to the average~$0.424 {\times} 10^{-4}$~emu/mol of two band-theoretical values without correlation~\cite{SI2012, LXH2013}. Then $\chi_\mathrm{orb} = \chi_0 - \chi_\mathrm{core}^\mathrm{dia} - \chi_P - \chi_L$. 

The results are given in Table~\ref{table:Tindepchi}. 
\begin{table} [ht]
\caption{\label{table:Tindepchi} Temperature-independent magnetic susceptibility of LOFBS from high-temperature magnetization, estimated core diamagnetic ($\chi_\mathrm{core}^\mathrm{dia}$), Pauli ($\chi_P$), Landau diamagnetic($\chi_L$), and orbital ($\chi_\mathrm{orb}$) contributions, and effective mass ratios~$m^*/m$. Susceptibility units are $10^{-4}$~emu/mol. See text for details.}
\begin{ruledtabular}
\begin{tabular}{cccccc}
$\chi_0$ & $\chi_\mathrm{core}^\mathrm{dia}$ & $\chi_P$ & $m^*/m$ & $\chi_L$ & $\chi_\mathrm{orb}$ \\
\colrule
$-0.461$ & $-1.942$\footnote{From Refs.~\cite{CBK77} and \cite{MBM70}.} & 0.347\footnote{$\gamma_\mathrm{spht}$ from Ref.~\cite{YHWCFM2013}.} & 0.82 & $-0.172$ & 1.306 \\
$''$ & $''$ & 0.270\footnote{$\gamma_\mathrm{spht}$ from Ref.~\cite{SLYZ14}.} & 0.64 & $-0.221$ & 1.432 \\
$''$ & $''$ & 0.453\footnote{$\gamma_\mathrm{spht}$ from Ref.~\cite{KTNT2017}.} & 1.36 & $-0.104$ & 1.010 
\end{tabular}
\end{ruledtabular}
\end{table}
The considerable variations in values of $\chi_P$ and related quantities are due to the variation in reported values of $\gamma_\mathrm{spht}$.

\subsubsection{\label{disc:Tdepmag} Temperature-dependent magnetization; models} 

For each field $M_0(H)$ and corrections for the trace Ce and Gd impurities were subtracted from $M_\mathrm{obs}(H,T)$ to obtain the intrinsic magnetization~$M_\mu(H,T)$ [Fig.~\ref{fig:parafit}(b)]. 
The parameter values resulting from the fits of the BFM and TLM to $M_\mu(H,T)$ are given in 
Table~\ref{table:CWfit}.
% revised
\begin{table} [ht]
\caption{\label{table:CWfit} Parameters from fits of the BFM and TLM [Eqs.~(\ref{eq:brillfit}) and (\ref{eq:twolevelfit}), respectively] to intrinsic magnetization $M_\mu(H,T)$ (Fig.~\ref{fig:parafit}).}
\begin{ruledtabular}
\begin{tabular}{lll}
model & BFM & TLM \\
\colrule
Concentration~$c$ (ppm) & 1033(164) & 1500(20) \\
Max.\ magnetization $M_\mathrm{sat}$ (emu/mol) & 26.0(4) & 25.4(2) \\
Effective moment $\mu_\mathrm{eff}$ ($\mu_B$) & 4.5(7) & 3.03(3) \\
$J$ & 1.0(4) & 1.52(2)\,\footnote{$J = \mu_\mathrm{eff}/g$.} \\
Spin-orbit/CEF ratio $\lambda/\Delta$ (TLM) & {-}{-} & $-0.51(2)$ \\
$g$ & 4.4(1.7)\,\footnote{$g = \mu_\mathrm{eff}/J$.} & 2\,\footnote{Fixed, cf.\ Ref.~\onlinecite{Kittel}.} \\
Weiss temperature $T_W$~(K) & 1.67(11) & 1.73(10) \\
\end{tabular}
\end{ruledtabular}
\end{table} 

\noindent The two models are discussed below.

\paragraph{\label{disc:BFM} BFM.} The Brillouin function describes the paramagnetism of ions with unfilled shells for which the ground state configuration is the lowest lying $2J+1$ multiplet. A fit with the BFM [Eq.~(\ref{eq:brillfit})] was made with $g$ factor~$g = \mu_\mathrm{eff}/J$, since the three parameters cannot all be left free. The resultant~$g$ value seems large. There is no evidence for magnetic order down to 2~K, and the origin of $T_W$ is not clear.

A BFM fit using fixed Gd$^{3+}$ ionic parameters $g = 2$, $J = S = 7/2$, $\mu_\mathrm{eff} = 7$ yields a concentration~$\sim$850~ppm, $\sim$30 times larger than the measured level of Gd impurities (Sec.~\ref{exp:sample}). The trace Ce in the sample also cannot account for $M_\mu(H,T)$. No other $d$ or $f$ elements are present, so that one is led to suspect the presence of local-moment defects associated with Bi $6p$ states.

% revised
\paragraph{\label{disc:TLM} TLM.} The TLM [Eq.~(\ref{eq:twolevelfit})] is based on the conjecture that the sources of local magnetic moments are Bi $6p$ ions in crystal defects, e.g., bond disorder as in Ref.~\onlinecite{AL2019}. The energy levels of these electrons can be modeled as resulting from spin-orbit coupling combined with the crystal electric field (CEF) produced by a lack of inversion symmetry at the defect site~\cite{[{See, e.g., }] LBZ09}. 

\begin{figure} [ht]
\includegraphics[clip = ,width = 0.4\textwidth]{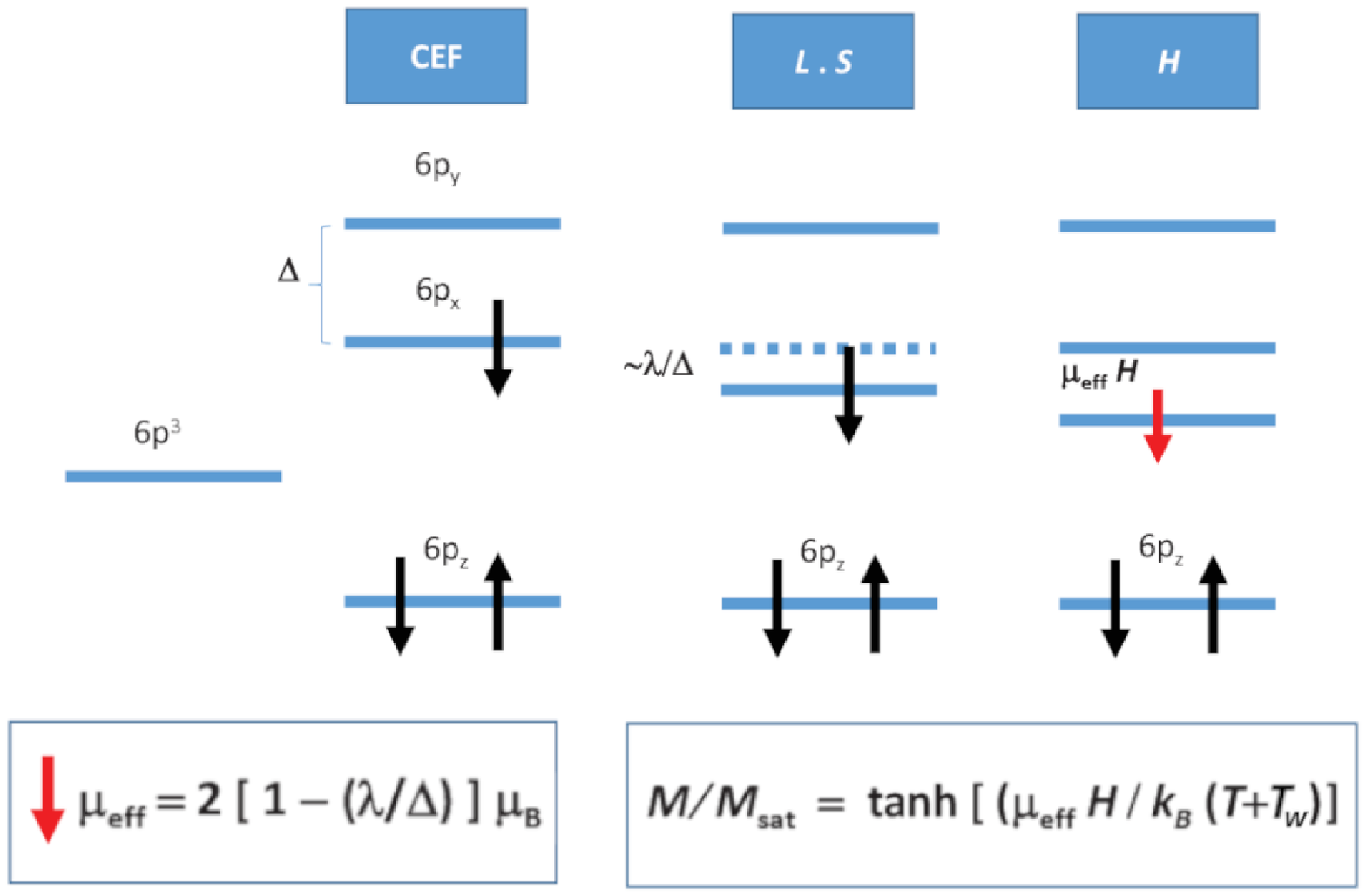}
\vspace{-0.5cm}\caption{\label{fig:SOC-Levels} Conjectured level scheme for the TLM.}
\end{figure}

Figure~\ref{fig:SOC-Levels} is a schematic depiction of the TLM\@. 
It has two parameters, the spin-orbit coupling~$\lambda$ and the CEF splitting~$\Delta$ between the $6p_x$ and $6p_y$ levels of the affected Bi atoms~\cite{Kittel}. The lowest-lying of the three resulting levels is fully occupied by 2 of the 3 $6p$ electrons. The other two levels are higher and differ in energy by $\Delta$, and the lower of these is half filled by the remaining electron. Spin-orbit coupling changes this level's energy in first order by an amount proportional to $\lambda/\Delta$, as depicted in Fig.~\ref{fig:SOC-Levels}. The level splits in a magnetic field by an amount proportional to the effective moment, which to first order is~$\mu_{\rm eff}=g[1-\lambda/\Delta]$~\cite{Kittel}. 

% revised
Of the two models, the TLM seems a more realistic description of Bi $6p$ magnetism, and the BFM-fit $g$ value of 4.4 is hard to reconcile with Bi defect moments. Analysis of $^{19}$F spin-lattice relaxation data using the TLM is given below in (Sec.~\ref{disc:slrelax}), but the results do not depend strongly on which model is used.
 
% revised
\paragraph{\label{par:CWansatz} Curie-Weiss \emph{ansatz}.} The Curie-Weiss phenomenology used in the BFM and TLM fits is only valid for small argument~$x$, and it is not surprising that there are systematic deviations for large $x$. If spin-spin interactions are involved the simplest scenario for taking them into account is the molecular-field approach, where interactions between spins are approximated by a single local field proportional to the system magnetization. The total field is then $H_0 + \Lambda M_\mu$, where $\Lambda$ is the Weiss molecular field constant. The argument~$x$ of the BFM and TLM then becomes $x = \mu_\mathrm{eff}(H_0 + \Lambda M_\mu)/k_BT$.

% revised
Fits using this \emph{ansatz} are shown in the Appendix. Agreement is not improved for large $x$ and, more importantly, the maximum molecular field~$\Lambda M_\mathrm{sat}$ is nearly 10~kOe. As discussed in Secs.~\ref{disc:spectra} and \ref{disc:defect} below, $^{19}$F NMR results indicate that this is four orders of magnitude larger than either dipolar or RKKY coupling fields. Curie-Weiss behavior can be due to CEF effects rather than spin-spin interactions~\cite{[{See, e.g., }] Bout73}, but a method of extending the Curie-Weiss phenomenology to large $x$ has not yet been found.

\subsection{\label{disc:spectra}\boldmath $^{19}$F Spectra}

\subsubsection{\label{disc:shift} $^{19}$F Frequency shifts}

The Knight shift resulting from conduction-band paramagnetism is the largest term in simple metals, and is typically in the 10$^3$-10$^4$~ppm range. The small shifts in LOFBS ($\lesssim 100$~ppm, Fig.~\ref{fig:KvsT}) suggest a low conduction-band density at the $^{19}$F site~\cite{SI2012, LXH2013}, consistent with the conduction channel being localized on the BiS$_2$ layers (Fig.~\ref{fig:struct}). 

In addition, the normal-state temperature dependence of the shift indicates coupling between $^{19}$F spins and a temperature-dependent contribution to the magnetism. An obvious candidate for this is the defect magnetism described above in Sec.~\ref{rslts:magresults}. 

\begin{figure} [ht]
\includegraphics[clip=,width=0.4\textwidth]{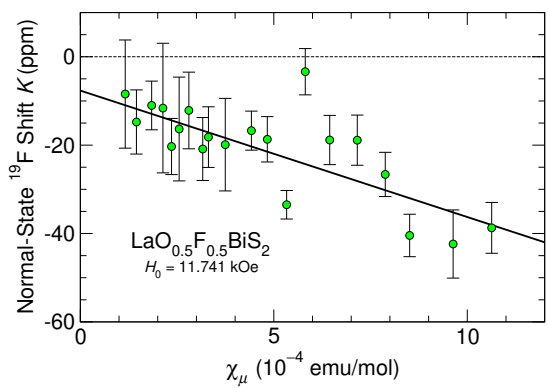}
\caption{\label{fig:Kvschi} Normal-state $^{19}$F frequency shift $K(T)$ vs defect susceptibility~$\chi_\mu(T)$ (Clogston-Jaccarino plot) in LOFBS powder, $T > T_c$. Line: least-squares fit to the data.}
\end{figure}

Figure~\ref{fig:Kvschi} is a Clogston-Jaccarino plot~\cite{CGJY64} of $K(T)$ data (Fig.~\ref{fig:KvsT}) vs $\chi_\mu(T)$ for $T > T_c$.
The linear fit yields intercept~$K_0$ and slope (coupling constant)~$A_K = dK/d\chi_\mu$. $K_0 = -7(3)$~ppm is the temperature-inde\-pendent sum of the conduction-band Knight shift and a contribution from core diamagnetism~\cite{CBK77}. These cannot be determined separately without further information. 

$A_K$ is also the slope~$A_{\Delta H} = d\Delta H/dM_\mu$ of the absolute field shift $\Delta H = H_0 K$ vs magnetization~$M_\mu$, since $d\Delta H/dM = dK/d\chi$. Thus the slope~$A_{\Delta\omega} = d\Delta\omega/dM_\mu$ of the NMR frequency shift~$\Delta\omega = \gamma_n\Delta H$ vs $M_\mu$ is $\gamma_n A_{\Delta H}$, where $\gamma_n$ is the nuclear gyromagnetic ratio. Fit values of these slopes are given in Table~\ref{tab:lineshape}.
\begin{table} [ht]
\caption{\label{tab:lineshape} Shift and linewidth coupling constants from Clogston-Jaccarino linear fits to shift and linewidth data. $^{19}\gamma$ is the $^{19}$F gyromagnetic ratio.} 
\begin{ruledtabular}
\begin{tabular}{cccc}
$A_K$, $A_{\Delta H}$ & $A_{\Delta\omega}$ & $A_{T^*_{2e}}^\mathrm{obs}$ & $A_{T^*_{2e}}^{\mathrm{calc}}$ \\
\hline
$-0.033(6)$ & $-0.83(15)$ & $2.8(2)$ & 1.94 \\
mol/emu & mol/ms-emu & mol/ms-emu & mol/ms-emu \\
\end{tabular}
\end{ruledtabular}
\end{table}

\subsubsection{\label{disc:linewidths} $^{19}$F Linewidths}

The nearly temperature-independent Gaussian linewidth $1/T^*_{2G}$ is due to nuclear dipole-dipole interactions, whereas the Lorentzian linewidth (exponential rate)~$1/T^*_{2e}$ arises from defect moments. There are two contributions to the latter~\cite{WaWa74}: defect-moment dipolar fields, and indirect RKKY interactions via conduction-electron polarization. 

The dipolar contribution in a free-electron metal is given by
\begin{eqnarray} 
1/T^{*\,\mathrm{dip}}_{2e} & = & \frac{8\pi^2}{9\sqrt{3}}N_0c\,\gamma_m \gamma_n \hbar \langle J_z \rangle \\
& = & A_{T_{2e}}^\mathrm{dip}M_\mu,\ A_{T_{2e}}^\mathrm{dip} =\frac{8\pi^2\gamma_n}{9\sqrt{3}V_\mathrm{mol}} \label{eq:invT2e} \,,
\end{eqnarray}
where $N_0$ is the formula-unit density, $c$ is the defect concentration, $\gamma_m$ is the defect-spin gyromagnetic ratio, $\langle J_z \rangle$ is the defect polarization, $M_\mu$ is the molar defect magnetization, and $V_\mathrm{mol} = 68.05~\text{cm}^3$/mol is the molar volume. 

The RKKY contribution is given by
\begin{eqnarray} 
1/T^{*\,\mathrm{RKKY}}_{2e} & = & \frac{4\pi}{3} A_{T_{2e}}^\mathrm{RKKY} \langle S_z \rangle \,, \text{ where} \\
A_{T_{2e}}^\mathrm{RKKY} & = & \frac{2\pi A_\mathrm{hf}N^2(E_F){\cal J}_\mathrm{ex} E_F}{(2k_F)^3} \,;
\end{eqnarray}
here $A_\mathrm{hf}$ is the $^{19}$F/conduction-band hyperfine coupling, $N(E_F)$ is the conduction-band density of states at the Fermi energy~$E_F$, ${\cal J}_\mathrm{ex}$ is the defect-moment/conduction-band exchange interaction, and $k_F$ is the Fermi wave vector. 

% revised
In Sec.~\ref{disc:slrelax} $N(E_F){\cal J}_\mathrm{ex}$ and an upper limit on $A_\mathrm{hf}$ are estimated from $^{19}$F spin-lattice relaxation data. The resulting upper limit on $1/T^{*\,\mathrm{RKKY}}_{2e}$ ($\langle S_z \rangle = S$) is $9.3(3) \times 10^{-3}~\text{ms}^{-1}$, which is three orders of magnitude smaller than experimental values~(Fig.~\ref{fig:echorates}). This is evidence that the dipolar interaction is dominant in LOFBS, which is assumed in the following.

\begin{figure} [ht] 
\includegraphics[clip = ,width = 0.4\textwidth]{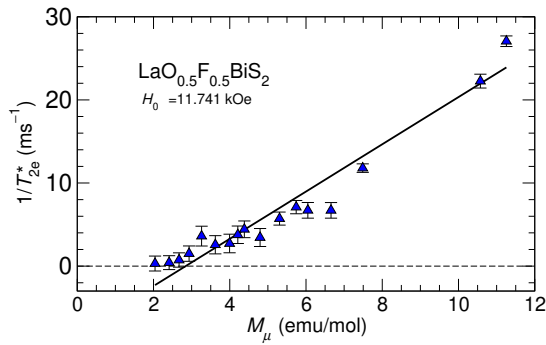}
\caption{\label{fig:T2starevsM} Exponential echo relaxation rate~$1/T^*_{2e} $ vs defect magnetization~$M_\mu$ in LOFBS powder. Line: linear fit to the data.}
\end{figure}

Figure~\ref{fig:T2starevsM} is a Clogston-Jaccarino plot of $1/T^*_{2e}$ vs $M_\mu$, showing the linear relation expected from Eq.~(\ref{eq:invT2e}).
Table~\ref{tab:lineshape} also gives the observed and calculated slopes~$A_{T^*_{2e}}$. The observed slope~$A_{T^*_{2e}}^{\mathrm{obs}}$ is in reasonable agreement with the calculated dipolar value~$A_{T^*_{2e}}^{\mathrm{calc}}$ [Eq.~(\ref{eq:invT2e})]. This indicates that the defect/$^{19}$F coupling is predominantly dipolar~\cite{WaWa74}. Dipolar coupling is also consistent with the observed spin-lattice relaxation in LOFBS, as discussed below in Sec.~\ref{disc:coupmech}. We note that $|A_{\Delta\omega}|$ is somewhat smaller than but comparable to~$A_{T^*_{2e}}^{\mathrm{obs}}$.

We conclude that $1/\gamma_mT^*_{2e}$ is the width of the Lorentzian defect dipole-field distribution, and that $\Delta H$ is a rough estimate of its mean~\footnote{The Gaussian component of the distribution that is due to coupling to neighboring nuclei has zero mean and would not contribute to the shift~\cite{Abra61}.}. We note that the mean dipolar shift vanishes for a randomly-oriented powder sample with an isotropic susceptibility, but this is unlikely to be the case in LOFBS.

The offset and corresponding negative value of the fit-line intercept are not well understood. They may be artifacts of the strong statistical correlation between $1/T^*_{2e}$ and $1/T^*_{2G}$ in fits of Eqs.~(\ref{eq:echoi}) and (\ref{eq:echoq}) (correlation coefficient $-0.894$); this propagates systematic error between the parameters. 

\subsection{\label{disc:slrelax} \boldmath $^{19}$F spin-lattice relaxation}

\subsubsection{\label{disc:Korringa} Korringa relaxation?} 

% revised
From the inset to Fig.~\ref{fig:pexprelax}(a) there is no evidence for the Korringa spin-lattice relaxation~$T_{1K}T = \text{const.}$ expected from direct $^{19}$F coupling to the conduction band or bands. The data suggest a rough lower limit on $T_{1K}T$ of ${\sim}300$~s-K, but it is unlikely to be this small given the large high-temperature slope of $\tau_1T$\@. Furthermore, Korringa relaxation is homogeneous and hence a simple exponential, but the power~$p$ of the observed stretched exponential shows no sign of approaching 1 at high temperature [Fig.~\ref{fig:pexprelax}(b)]. Introducing a simple exponential to the relaxation fit function worsens the least-squares fit [Fig.~\ref{fig:diffplot}(b)].

For a free-electron gas the Korringa product~${\cal S}_0 = K_K^2T_{1K}T = (\hbar/4\pi k_B)(\gamma_e/\gamma_n)^2$ is a constant; here $K_K$ is the conduction-band Knight shift and $\gamma_e$ and $\gamma_n$ are the electron and nuclear gyromagnetic ratios, respectively. For $\gamma_n = \gamma(^{19}\mathrm{F})$ ${\cal S}_0 = 2.973 {\times} 10^{-7}$~s-K, which for $T_{1K}T \gtrsim 300$~s~K yields a free-electron estimate of a upper bound~$|K_K| < 32(2)$~ppm on the Knight shift. This lies between the negative intercept~$K_0 = -7$~ppm of the Clogston-Jaccarino plot (Fig.~\ref{fig:Kvschi}) and the observed full shift~$K(T{\to}\infty) - K(T{=}0)$ of roughly 70~ppm (Fig.~\ref{fig:KvsT}) and is therefore consistent with the data, although the diamagnetic contribution is unknown (Sec.~\ref{disc:shift}). A value of $\cal S$ for LOFBS much larger than ${\cal S}_0$ seems unlikely; in non-transition elemental metals, e.g., $K^2T_{1K}T/{\cal S} \lesssim 2$~\cite{CBK77}. 

% revised
This result can be used to obtain an estimated upper limit on the $^{19}$F/conduction-band hyperfine field, given by
\begin{equation} \label{eq:Hhf}
H_\mathrm{hf} = K_K/\mu_BN(E_F)
\end{equation} 
in the free-electron approximation~\cite{CBK77}. With the upper bound on $|K_K|$ and the band-theoretic density of states~$N(E_F) = 1.31~\text{eV}^{-1}/$formula unit~\cite{SI2012, LXH2013}, $H_\mathrm{hf} < 4.2~\text{kOe}/\mu_B$ from Eq.~(\ref{eq:Hhf}). Lattice sums yield dipolar fields from Bi moments at F sites that are an order of magnitude smaller than this. For comparison, $N(E_F) = 0.47~\text{eV}^{-1}/$formula unit~\cite{SI2012, LXH2013} and $H_\mathrm{hf} < 180~\text{kOe}/\mu_B$ for Gd$^{3+}$ impurities in the 3D metal~La$_{1-c}$Gd$_c$Al$_2$~\cite{MSW72, SiWe73}. 

\subsubsection{\label{disc:coupmech} Coupling mechanisms: LD, TD, BGS}

\begin{table*} [ht]
\caption{\label{table:invtau1max} Experimental and calculated spin-lattice relaxation coupling coefficients in LOFBS, $H_0 = 11.74$~kOe. \\
\emph{ Observed}: Maximum rate~$1/\tau_{1\,\mathrm{max}} = 5.23(5)$~s$^{-1}$ (Fig.~\ref{fig:pexprelax}), temperature~ of maximum~$T_\mathrm{max} = 8.316$~K. \\
\emph{Calculated}: correlation time~$\tau_{c}(T_\mathrm{max}) = 1/\omega_i$, coupling constant~$|\Gamma|^2_\mathrm{exp} = 2\omega_i/\tau_{1\,\mathrm{max}}$, $|\Gamma|^2_\mathrm{calc}$ (Sec.~\ref{disc:coupmech}), maximum static exponential relaxation rate~$T^{*-1}_{2e}(\text{max}) = 7.2(6){\times}10^4~\text{s}^{-1}$.}
\begin{ruledtabular}
\begin{tabular}{lccccccc}
Mechanism & $\omega_i$ (s$^{-1}$) & $\tau_c(T_\mathrm{max})$ (s) & $|\Gamma|^2_\mathrm{exp}$ (s$^{-2}$) & $|\Gamma|_\mathrm{exp}$ (s$^{-1}$) & $|\Gamma|_\mathrm{calc}$ (s$^{-1}$) & $|\Gamma|_\mathrm{calc}/|\Gamma|_\mathrm{exp}$ & $|\Gamma|_\mathrm{exp}/T^{*-1}_{2e}(\text{max})$ \\
\colrule
LD ($\omega_i {=} \omega_n$) & $2.96 {\times} 10^8$ & $3.38{\times} 10^{-9}$ & $3.09(3) {\times} 10^9$ & $5.56(3) {\times} 10^4$ & $6.21 {\times} 10^4$\,\footnote{From Eq.~(\ref{eq:C2LD}).} & 1.117(6) & 0.77(7) \\
TD ($\omega_i {=} \omega_m$) & $2.55 {\times} 10^{11}$ & $3.93 {\times} 10^{-12}$ & $2.66(3) {\times} 10^{12}$ & $1.631(8) {\times} 10^6$ & $9.49 {\times} 10^4$\,\footnote{From Eq.~(\ref{eq:C2TD}).} & 0.06 & 23 \\ 
BGS ($\omega_i {=} \omega_m$) & $''$ & $''$ & $''$ & $''$ & $< 32.9$\,\footnote{Upper limit from Eq.~(\ref{eq:C2BGS})using $N(E_F){\cal J}_\mathrm{ex}$ from Eq.~(\ref{eq:taucT}).} & $ < 2{\times}10^{-5}$ & $''$ \\ 
\end{tabular}
\end{ruledtabular}
\end{table*}

McHenry, Silbernagel, and Wernick~\cite{[{}][{, and references therein.}] MSW72} (MSW) reviewed potential mechanisms for host nuclear relaxation by paramagnetic impurities in metals. Their data fit the root-exponential relaxation function~$\exp[-(t/\tau_1)^{1/2}]$ expected from dipolar relaxation in the dilute limit~\cite{TsHa68}. Stretched-exponential relaxation~$\exp\left[-(t/\tau_1)^p\right]$ is also observed in LOFBS, albeit with a larger power~$p \approx 0.7$--0.8 (Sec.~\ref{rslts:slrelax}). Relations appropriate for $p = 1/2$ are used in the following, with the understanding that the results are only qualitative. 

Mechanisms for host nuclear relaxation by paramagnetic impurities~\cite{GPGH1971, MSW72} include coupling via dipolar fields to longitudinal and transverse defect-spin fluctuations (LD and TD respectively), RKKY coupling [Benoit-de Gennes-Silhouette (BGS)~\cite{BGS1963}], and coupling to virtual defect-spin excitations (Giovannini-Heeger (GH)~\cite{GH1969}). 

For the LD, TD, and BGS mechanisms the root-exponential rate~$1/\tau_1$ at high temperatures is related to a defect-spin correlation time~$\tau_c$ by~\cite{LT1968, GPGH1971} \footnote{Sometimes referred to as Bloembergen-Purcell-Pound (BPP) or Redfield relaxation.}
\begin{equation} \label{eq:invtau1Lor}
\frac{1}{\tau_1(\tau_c)} = |\Gamma|^2\frac{\tau_c}{1 + (\omega_i \tau_c)^2} \,. 
\end{equation} 
Here $|\Gamma|^2$ is the nuclear-spin/defect-spin coupling coefficient, and $\omega_i = \gamma_i H_0$, with $i = n$ (nuclear) for the LD mechanism and $i = m$ (magnetic defect) for the TD and BGS mechanisms. The observed $^{19}$F spin-lattice relaxation field and temperature dependencies in LOFBS (Fig.~\ref{fig:pexprelax}) strongly suggest this behavior, with $\tau_c$ a monotonically decreasing function of increasing temperature. (The GH mechanism yields a temperature- and field-independent rate apart from the saturation effect discussed below, and is not considered further.) 

From Eq.~(\ref{eq:invtau1Lor}) $1/\tau_1(\tau_c)$ exhibits a maximum~$1/\tau_{1\,\mathrm{max}}$ for $\omega_i\tau_c = 1$, so that $|\Gamma|^2 = 2\omega_i/\tau_{1\,\mathrm{max}}$. These differ for TD/BGS and LD by the large factor~$\omega_m/\omega_n$. 

Calculated values $|\Gamma|^2_\mathrm{calc}$ at high temperatures for the three mechanisms in the free-electron approximation are~\cite{MSW72} 
\begin{eqnarray}
\text{LD:}\ |\Gamma|^2_\mathrm{calc} & = & 16\pi^3(\gamma_n M_\mathrm{sat})^2\textstyle{\frac{1}{3}}(J+1)/J \nonumber \\
& & \times \overline{\sin^2\theta\cos^2\theta} \,, \label{eq:C2LD} \\
\text{TD:}\ |\Gamma|^2_\mathrm{calc} & = & \frac{8\pi^3}{9}(\gamma_n M_\mathrm{sat})^2\textstyle{\frac{1}{3}}(J+1)/J \nonumber \\ 
& & \times [\overline{(1 - 3\cos^2\theta)^2 + 9\sin^4\theta}] \,, \label{eq:C2TD} \\
\text{BGS:}\ |\Gamma|^2_\mathrm{calc} & = & \frac{16\pi^3}{9}\left(\frac{\pi A {\cal J}_\mathrm{ex} N^2(E_F)E_F N_0c}{2\hbar k_F^3}\right)^{\!\!2} \nonumber \\
& & \times\textstyle{\frac{1}{3}}J(J+1) \,, \label{eq:C2BGS} 
\end{eqnarray}
where $M_\mathrm{sat} = N_0c\,\mu_\mathrm{eff}$, and the overlines are averages over angles~$\theta$ to defect sites. Experimental and calculated values of $|\Gamma|^2$ for these mechanisms are given in Table~\ref{table:invtau1max}. Their comparison yields strong evidence that the LD mechanism is dominant in LOFBS\@. 

% revised
There is reasonable agreement between calculated and experimental values for the LD results ($|\Gamma|^2_\mathrm{calc}/|\Gamma|^2_\mathrm{exp} \approx 1$), but there is a large discrepancy for the TD results: $|\Gamma|^2_\mathrm{calc}/|\Gamma|^2_\mathrm{exp} \ll 1$. In addition, the ratio~$|\Gamma|_\mathrm{exp}/T^{*-1}_{2e}(\text{max})$, where $T^{*-1}_{2e}(\text{max}) = A_{T^*_{2e}}M_\mathrm{sat}$ is the maximum exponential static relaxation rate (Sec.~\ref{disc:linewidths}), is also of order unity for the LD mechanism; this would be expected because the two couplings were of the same origin. For the TD mechanism this ratio is $\gg 1$, which is again evidence against it.
 
The parameters in Eq.~(\ref{eq:C2BGS}) for the BGS mechanism are defined above in Sec.~\ref{disc:linewidths}. From the upper bound on $H_\mathrm{hf}$ derived above in Sec.~\ref{disc:Korringa}, $A < 5.6{\times}10^7~\text{s}^{-1}$. Then Eq.~(\ref{eq:C2BGS}) yields an upper limit for the BGS $|\Gamma|^2_\mathrm{calc}$ that is very much smaller than the experimental value (Table~\ref{table:invtau1max}) and can be neglected. 

\subsubsection{\label{disc:sateffs} Saturation effects}

Equation~(\ref{eq:invtau1Lor}) does not include the effect of saturation of defect moments at low temperatures and high fields, which leads to a reduction of $1/\tau_1$~\cite{MSW72}. We assume the TLM for analysis of this effect, since as discussed above we believe it to be the more appropriate model. A BFM analysis (not shown) produces nearly the same numerical results.

For the LD mechanism the modification of Eq.~(\ref{eq:invtau1Lor}) is of the form~\cite{BGS1963, MSW72}
\begin{equation} \label{eq:invtau10}
\frac{1}{\tau_1} = \frac{1}{\tau_1^0}\frac{d\tanh(x)}{dx} \,,
\end{equation}
so that
\begin{equation} \label{eq:invtau10Lor}
\frac{1}{\tau_1^0(T)} = |\Gamma_0(H_0,T)|^2 \frac{\tau_c(T)}{1 + [\omega_i\tau_c(T)]^2} \,, 
\end{equation}
with
\begin{equation} \label{eq:Aprime2}
|\Gamma_0(H_0,T)|^2 = \frac{|\Gamma|^2}{d\tanh(x)/dx} \,.
\end{equation} 

\begin{figure} [ht]
\includegraphics*[clip = ,width = 0.4\textwidth]{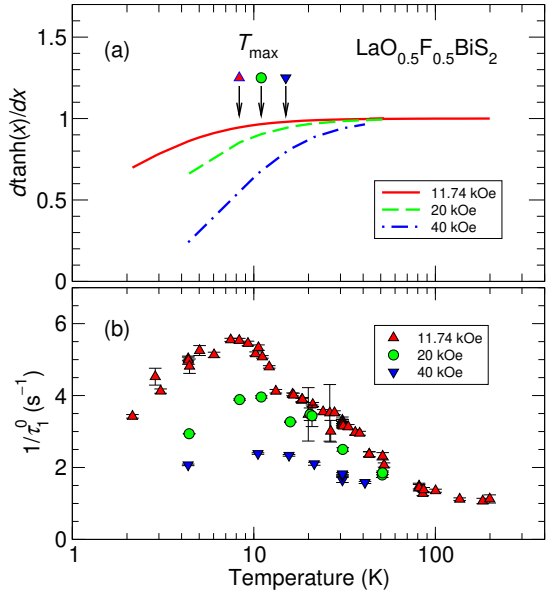}
\caption{\label{fig:invtau10vsT} Field and temperature dependencies of factors in Eq.~(\ref{eq:invtau10}). (a)~Saturation coefficient~$d\tanh(x)/dx$. Symbols: temperatures~$T_\mathrm{max}$ where $1/\tau_1$ is maximum. (b)~Relaxation rate~$1/\tau_1^0$.}
\end{figure}

Figure~\ref{fig:invtau10vsT}(a) gives $d\tanh(x)/dx$ using the Curie-Weiss form~$x = \mu_\mathrm{eff}H/k_B(T + T_W)$ with $T_W = 1.7$~K from fits to the magnetization data (Table~\ref{table:CWfit}). 
Although the reduction of $\tanh(x)/dx$ at and below $T_\mathrm{max}$ is significant for $H_0= 20$~kOe and especially $40$~kOe [Fig.~\ref{fig:invtau10vsT}(a)], for $H_0= 11.74$~kOe $1/\tau_1^0(T_\mathrm{max})$ from Eq.~(\ref{eq:invtau10}) is modified by less than 10\% and the conclusions of Sec.~\ref{disc:coupmech} above are not affected.

The field and temperature dependencies of $1/\tau_1^0$ obtained from $1/\tau_1$ and Eq.~(\ref{eq:invtau10}) are shown in Fig.~\ref{fig:invtau10vsT}(b). 

\subsubsection{\label{disc:defect} Defect-spin relaxation and exchange coupling} 

\begin{figure} [ht]
\includegraphics[clip =, width = 0.4\textwidth]{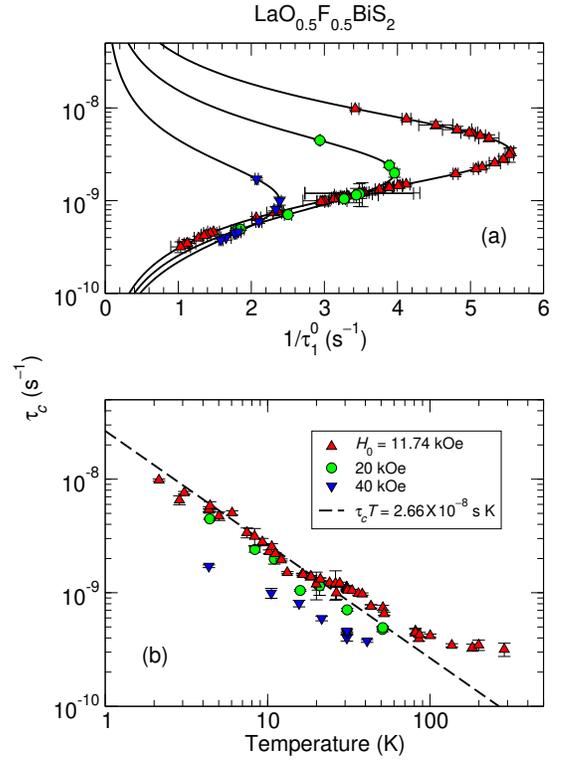} 
\caption{\label{fig:taucondtanhxdx} (a)~defect-spin correlation times~$\tau_c$ obtained from $1/\tau_1^0$ and Eq.~(\ref{eq:invtau10}). Curves: Eq.~(\ref{eq:invtau10Lor}). (b)~Field and temperature dependencies of $\tau_c$. Dashed line: fit of $\tau_cT = \text{const.}$ to 11.74 kOe data.}
\end{figure}

\paragraph{\label{sec:taucT} Field and temperature dependencies of $\tau_c$.} These have been determined by solving Eq.~(\ref{eq:invtau10Lor}) for $\tau_c(H,T)$ point by point from $\tau_1^0(H,T)$ [Fig.~\ref{fig:invtau10vsT}(b)) with $|\Gamma_0|^2$ from Eq.~(\ref{eq:Aprime2}). The results are shown in Fig.~\ref{fig:taucondtanhxdx}(a).
The curves are solutions of Eq.~(\ref{eq:invtau10Lor}), not fits; the equation is quadratic in $\tau_c$ and its two solutions are on opposite sides of the maximum in $1/\tau_1^0(T)$. 

The results for $\tau_c(H,T)$ obtained in this way are shown in Fig.~\ref{fig:taucondtanhxdx}(b). For $H_0 = 11.74$~kOe the data are roughly consistent below $\sim$100~K with $\tau_c T = 2.58(7) {\times} 10^{-8}$~s-K\@. 
% revised
For $H_0 = 20$ and $40$~kOe $\tau_c(T)$ is smaller than for $11.74$~kOe but with essentially the same slope. This is after correction for the reduction of $d\tanh(x)/dx$ [Fig.~\ref{fig:invtau10vsT}(b)], and thus is an intrinsic field dependence. It is not well understood.

% revised
This result justifies using the Korringa expression~\cite{Tayl75, Barn81}
\begin{equation} \label{eq:taucT}
1/\tau_cT = 4\pi(k_B/\hbar)[N(E_F){\cal J}_\mathrm{ex}]^2 \,,
\end{equation}
where ${\cal J}_\mathrm{ex}$ is the defect/conduction-electron exchange coupling.Table~\ref{tab:pars} gives $N(E_F){\cal J}_\mathrm{ex}$ using the observed value of $\tau_cT$ and Eq.~(\ref{eq:taucT}), and ${\cal J}_\mathrm{ex}$ using the average~$N(E_F) = 1.31~\text{eV}^{-1}/$formula unit from band theory~\cite{SI2012, LXH2013}. 
%\vspace{-10pt}
\begin{table} [ht]
\caption{\label{tab:pars} Defect/conduction-band DOS-spin exchange product~$N(E_F){\cal J}_\mathrm{ex}$ from Eq.~(\ref{eq:taucT}), exchange constant~${\cal J}_\mathrm{ex}$ using $N(E_F)$ from Refs.~\onlinecite{SI2012} and \onlinecite{LXH2013}, and defect RKKY correlation times~$\tau_m^\mathrm{RKKY}$ from Eq.~(\ref{eq:RKKY}) for LOFBS\@. Values for Gd-doped LaAl$_2$ (Ref.~\onlinecite{MSW72}) are included for comparison.} 
\begin{ruledtabular}
\begin{tabular}{lccc}
 & $N(E_F){\cal J}_\mathrm{ex}$ & ${\cal J}_\mathrm{ex}$ (meV) & $\tau_m^\mathrm{RKKY}$ (s) \\ 
\hline
LOFBS & 0.0047 & 3.6 & $2.3 \times 10^{-7}$ \\
La$_{1-c}$Gd$_c$Al$_2$ & 0.09 & 90 & $6.4 \times 10^{-10}$
\end{tabular}
\end{ruledtabular}
\end{table}
For comparison, $N(E_F){\cal J}_\mathrm{ex}$ and ${\cal J}_\mathrm{ex}$ for dilute Gd impurities in the 3D compound~La$_{1-c}$Gd$_c$Al$_2$~\cite{MSW72} are also given.

These results can be used to estimate the effect of the defect spins on the superconductivity of LOFBS\@. From the Abrikosov-Gor'kov theory~\cite{AbGo61, MuZi70}, the suppression~$\Delta T_c$ of $T_c$ by a concentration~$c$ of spin-$S$ impurities is given by
\begin{equation}
k_B\Delta T_c = -\frac{\pi^2}{8} c N(E_F){\cal J}_\mathrm{ex}^2 \,,
\end{equation}
which from the above results yields $\Delta T_c \approx -0.18$~mK\@. The effect is negligible. 

\paragraph{\label{disc: defect} Defect-defect interactions.} Coupling between dilute defect spins will also affect their fluctuation spectra. For example, RKKY Gd-Gd interactions are observed to dominate Gd$^{3+}$ spin fluctuations in La$_{1-c}$Gd$_c$Al$_2$~\cite{MSW72}. 

A rough estimate of the defect-spin fluctuation rate due to dipolar coupling is given by
\begin{equation} \label{eq:dipolar}
1/\tau_m^\mathrm{dip} \approx \gamma_m\mu_\mathrm{eff}/R^3 \,,
\end{equation}
where we take $\gamma_m = \mu_\mathrm{eff}/\hbar J$ using the average~$\mu_\mathrm{eff} \approx 3.7~\mu_B$ of the two model values (Table~\ref{table:CWfit}) for the defect moment, and $R = (3/4\pi N_0c)^{1/3}$ is the mean spacing between local moments. This yields $\tau_c^\mathrm{dip} \approx 6 {\times} 10^{-7}$~s, considerably longer than the observed values [Fig.~\ref{fig:taucondtanhxdx}(b)]. 

In a free-electron picture the defect RKKY correlation time~$\tau_m^\mathrm{RKKY}$ is given by~\cite{MSW72}
\begin{equation} \label{eq:RKKY}
1/\tau_m^\mathrm{RKKY} = \textstyle \frac{1}{9}\, [\frac{1}{6}\pi S(S+1)]^{1/2}N(E_F){\cal J}_\mathrm{ex}^{\,2} c/\hbar \,.
\end{equation}
Values of $\tau_m^\mathrm{RKKY}$ for LOFBS and La$_{1-c}$Gd$_c$Al$_2$ from Eq.~(\ref{eq:RKKY}) are also given in Table~\ref{tab:pars}. For LOFBS $\tau_m^\mathrm{RKKY}$ is again considerably longer than observed. We conclude that defect-defect interactions are not important for defect-spin relaxation, so that defect-spin fluctuations are due to Korringa mechanism by conduction electrons.

% revised
These results also determine the order of magnitudes~$1/\gamma_m^i\tau_m$, $i = \text{dip}$ and RKKY, of defect spin-spin interaction fields. They are both a gauss or less, compared to the nearly 10~kOe needed for molecular-field fits to the magnetization data (Sec.~\ref{par:CWansatz} and the Appendix).

\section{SUMMARY AND CONCLUSIONS} \label{sec:concl}~

We have carried out a magnetization and $^{19}$F NMR study of ambient-pressure-grown LaO$_{0.5}$F$_{0.5}$BiS$_2$ (LOFBS). The measurements reveal dilute local magnetic moments of a few $\mu_B$, the concentration of which ($\sim$1000~ppm) is an order of magnitude greater than the measured levels of $4f$ impurities. Thus the local moments are associated with structural defects, most likely magnetic Bi $6p$ states in the BiS$_2$ layers that are decoupled from the conduction band by defects.

% revised
Either of two models describe the defect-moment magnetism in LOFBS: a Brillouin function model (BFM), appropriate for weakly interacting spins in a paramagnet, or a two-level scenario (TLM), in which Bi $6p$ states acquire effective moments determined by a combination of crystal electric fields and spin-orbit coupling. The TLM fits cannot determine these separately, only their ratio. Both models fit the data, but both require a Curie-Weiss argument~$x = \mu_\mathrm{eff}H_0/k_B(T + T_W)$. The value of the Weiss temperature~$T_W$ is close to 1.7~K for both models, but both fits deviate from the data for large $x$ where this form is not justified. A molecular-field approximation~$x = \mu_\mathrm{eff}(H_0 + \Lambda M_\mu/k_BT$ also provides good fits, but with a molecular-field constant~$\Lambda$ far too large to be compatible with limits on defect spin-spin interactions from the $^{19}$F NMR data.

% revised
The NMR frequency shift, linewidth, and spin-lattice relaxation are consistent with a predominantly dipolar coupling between $^{19}$F nuclei and defect moments. Any direct conduction-band Korringa contribution to the observed $^{19}$F spin-lattice relaxation is very small, so that it is therefore dominated by defect-moment fluctuations. The defect-moment fluctuation rates~$1/\tau_c$ derived from $^{19}$F relaxation data exhibit Korringa relaxation $\tau_cT = \text{const.}$ due to the conduction band. This is inconsistent with a picture in which the defect moments are substitutional at La sites in the blocking layers (cf.\ Fig.~\ref{fig:struct}) as would be the case for, e.g., extrinsic $4f$ ions. The Korringa law for the defect-spin relaxation rate involves the conduction band, and we tend to favor the TLM as a more appropriate picture of $6p$ moments in defects.
 
The situation in LOFBS may be similar to the paramagnetism that has been detected in other Bismuth compounds, particularly $\alpha$-Bi$_2$O$_3$ where paramagnetism was observed at a similar concentration level~\cite{KOS06}. 

More experiments are needed. Magnetization and NMR studies of pressure-grown and as-grown samples and studies of samples under pressure would be useful to determine the behavior of defect spin dynamics with changes of superconducting coherence and the attainment of 
higher $T_c$s in this family of materials.

%\newpage 
\begin{acknowledgments}

% \vspace{-10pt} 
This work was supported in part by grants from the National Science Foundation, HRD/CREST-2112554 and DMR/PREM-1523588, the University of California, Riverside Academic Senate, and 
the National Natural Science Foundations of China, No. 12174065.
Research at the University of California, San Diego was supported by the US Department of Energy, Basic Energy Sciences, under Grant DE-FG02-04ER46105. 
\end{acknowledgments}

%\vspace{-10pt} 
\appendix* \section{Molecular-Field BFM and TLM Fits to the Magnetization}

\begin{figure} [ht]
\includegraphics[clip = ,width = 0.4\textwidth]{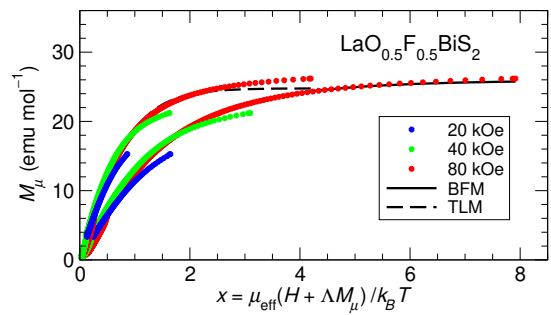}
\caption{\label{fig:molfield} Fits to the intrinsic magnetization~$M_\mu(H,T)$ of the BFM (solid curve) and the TLM (dashed curve) using the molecular field \emph{ansatz}, i.e., arguments $x = \mu_\mathrm{eff}(H + \Lambda M_\mu)/k_BT$\@. Cf.\ Sec.~\ref{par:CWansatz}.}
\end{figure}

\vspace{-10pt} Figure~\ref{fig:molfield} shows the results of fits of the two models to the data assuming a molecular field, i.e., for the argument~$x = \mu_\mathrm{eff}(H_0 + \Lambda M_\mu)/k_BT$\@. Table~\ref{table:molfldfit} gives the best-fit parameters.
\begin{table} [ht]
\caption{\label{table:molfldfit} Parameters from molecular-field fits of the BFM and TLM [Eqs.~(\ref{eq:brillfit}) and (\ref{eq:twolevelfit}), respectively] to intrinsic magnetization $M_\mu(H,T)$ (Fig.~\ref{fig:molfield}).}
\begin{ruledtabular}
\begin{tabular}{lcc}
Model & BFM & TLM \\
\colrule
Concentration (ppm) & 795(71) & 1424(18) \\
Saturation magnetization $M_\mathrm{sat}$ (emu/mol) & 26.1(5) & 25.5(2) \\
Effective moment $\mu_\mathrm{eff}$ ($\mu_B$) & 5.8(5) & 3.11(3) \\
$J$ (BFM) & 1.8(5) & {-}{-} \\
Spin-orbit/CEF ratio $\lambda/\Delta$ (TLM) & {-}{-} & $-0.56(2)$ \\
$g$ & 3.2(9)\,\footnote{$g = \mu_\mathrm{eff}/J$.} & 2\,\footnote{Fixed, cf.\ Ref.~\onlinecite{Kittel}.} \\
Mol.\ field const.~$\Lambda$ (mol-Oe/emu) & $-347(18)$ & $-363(17)$ \\
Max.\ mol.\ field~$\Lambda M_\mathrm{sat}$ (kOe) & $-9.0(4)$ & $-9.0(5)$ \\
\end{tabular}
\end{ruledtabular}
\end{table} 
Compared with the Curie-Weiss BFM and TLM fit values (Table~\ref{table:CWfit}), the parameters~$M_\mathrm{sat}$, $\mu_\mathrm{eff}$, BFM $J$, TLM $\lambda/\Delta$, and $g$ differ by no more than a factor of 2. The unphysically large values of the molecular field constant~$\Lambda$ are discussed in Sec.~\ref{par:CWansatz}.
\bibliography{LaOFBiS2}

\end{document}